# AI chatbots versus human healthcare professionals: a systematic review and meta-analysis of empathy in patient care

Alastair Howcroft[1,2,*] , Amber Bennett-Weston[2] , Ahmad Khan[3] , Joseff Griffiths[4] , Simon Gay[3] , Jeremy Howick[2]

[1]School of Computer Science, University of Nottingham, Jubilee Campus, Wollaton Road, Nottingham, Nottinghamshire NG8 1BB, United Kingdom
[2]Stoneygate Centre for Empathic Healthcare, Leicester Medical School, University of Leicester, George Davies Centre, Lancaster Road, Leicester, Leicestershire LE1 7HA, United Kingdom
[3]Leicester Medical School, University of Leicester, George Davies Centre, Lancaster Road, Leicester, Leicestershire LE1 7HA, United Kingdom
[4]School of Medicine, University of Nottingham, Queen's Medical Centre, Derby Road, Nottingham, Nottinghamshire NG7 2UH, United Kingdom

*Corresponding author. School of Computer Science, University of Nottingham, Jubilee Campus, Wollaton Road, Nottingham NG8 1BB, UK. E-mail: psyah9@nottingham.ac.uk

## Abstract

**Background:** Empathy is widely recognized for improving patient outcomes, including reduced pain and anxiety and improved satisfaction, and its absence can cause harm. Meanwhile, use of artificial intelligence (AI)–based chatbots in healthcare is rapidly expanding, with one in five general practitioners using generative AI to assist with tasks such as writing letters. Some studies suggest AI chatbots can outperform human healthcare professionals (HCPs) in empathy, though findings are mixed and lack synthesis.
**Sources of data:** We searched multiple databases for studies comparing AI chatbots using large language models with human HCPs on empathy measures. We assessed risk of bias with ROBINS-I and synthesized findings using random-effects meta-analysis where feasible, whilst avoiding double counting.
**Areas of agreement:** We identified 15 studies (2023–2024). Thirteen studies reported statistically significantly higher empathy ratings for AI, with only two studies situated in dermatology favouring human responses. Of the 15 studies, 13 provided extractable data and were suitable for pooling. Meta-analysis of those 13 studies, all utilising ChatGPT-3.5/4, showed a standardized mean difference of 0.87 (95% CI, 0.54–1.20) favouring AI ($P < .00001$), roughly equivalent to a two-point increase on a 10-point scale.
**Areas of controversy:** Studies relied on text-based assessments that overlook non-verbal cues and evaluated empathy through proxy raters.
**Growing points:** Our findings indicate that, in text-only scenarios, AI chatbots are frequently perceived as more empathic than human HCPs.
**Areas timely for developing research:** Future research should validate these findings with direct patient evaluations and assess whether emerging voice-enabled AI systems can deliver similar empathic advantages.

## Introduction

Empathic healthcare is well recognized for its positive impact on patient quality of life, satisfaction with care, and reduction of pain and psychological distress [1]. With recent advances in artificial intelligence (AI) technologies, chatbots are increasingly being integrated into patient care, even replacing human practitioner roles at times. For example, Wysa—a digital therapist—has been used by over 117 000 patients across 31 NHS Talking Therapy services, according to Wysa's official website [2]. These AI systems can interact with patients through text or speech, providing information, monitoring symptoms, offering support, and fulfilling other roles historically provided by human healthcare professionals (HCPs) [3]. It has also been reported that 20% of UK general practitioners (GPs) now use generative AI like ChatGPT for tasks such as assistance with writing patient correspondence [4]. However, despite the growing use of these technologies [5], there are concerns about whether AI chatbots can be as empathic as human HCPs [6], and thus leverage the benefits of empathic care for patients [7]. Such doubts align with the 2019 Topol Review, a UK government-commissioned roadmap for NHS technology, which concluded that 'empathy and compassion' remain 'essential human skill [s] that AI cannot replicate' [8].

Whilst individual studies have explored the potential of AI technologies to display empathic behaviours [9], findings are heterogeneous. Some report that AI bots can produce responses with higher levels of perceived empathy [10], whilst others cite issues suggesting the contrary, maintaining that they lack the warmth, nuanced understanding, and ability to form deep emotional bonds inherent in human interactions [11]. Existing literature reviews on the use of AI bots have primarily focused on disease diagnosis, clinical efficiency, ethical considerations, and their integration into healthcare systems [5, 12–14] and factual accuracy of medical advice compared to humans [15].

The lack of a synthesis of studies comparing AI technology with human empathy represents a significant gap in the literature. Such a synthesis can pave the way for more informed decisions about the integration of AI technologies into patient care, ensuring that such technological advancements contribute to (and don't detract from) the positive patient outcomes historically associated with empathic healthcare delivered by humans.

## Methods

The review is reported according to the PRISMA 2020 Checklist.

### Eligibility criteria

Studies were eligible if they empirically compared empathy between AI chatbots and human HCPs. Eligible study designs included quantitative or qualitative studies involving real patients, healthcare users, or authentic patient-generated data, such as emails, portal messages, or public forum posts. We included participants of any age, gender, or demographic engaging in formal or informal healthcare interactions. AI interventions were restricted to conversational agents using Large Language Models (LLMs), such as GPT-3/4, Claude, or Gemini, capable of unscripted dialogue.

Exclusions included healthy volunteers without healthcare needs and interactions outside healthcare contexts (e.g. education, customer service). Hypothetical patient scenarios (e.g. researcher-generated questions without a real patient source) [16] were excluded because they are further removed from real-world practice and may not capture the authentic phrasing or context of genuine healthcare interactions. Our aim was to include only interactions that, while varied in setting, still originated from actual patient-generated content. We also excluded studies utilising rule-based or scripted AI systems, opinion pieces, theoretical discussions, editorials, commentaries, and reviews lacking original empirical data.

### Information sources

We searched PubMed, Cochrane Library, Embase, PsycINFO, CINAHL, Scopus, and IEEE Xplore from inception to 11 November 2024. ClinicalTrials.gov, ICTRP, and ISRCTN were searched for completed studies with published results. Reference lists of included studies and relevant reviews were screened, and grey literature was searched via Google Scholar on 12 November 2024.

### Search strategy

The search strategy focused on three key groups of terms: empathy, AI technologies, and HCPs. Relevant keywords and index terms were identified through preliminary searches in IEEE Xplore and PubMed, supported by consultation with an academic librarian. Broad terms for AI and HCPs captured diverse technologies and roles, whereas empathy terms ('empathy', 'empathic', 'empathetic', 'compassion', and derivatives) were narrower to maintain specificity. Groups were combined with 'OR' within categories and 'AND' across categories. The full strategy is detailed in Appendix A.

### Selection process

The primary reviewer independently screened all titles and abstracts to ascertain which studies to include or exclude based on the predefined eligibility criteria. The abstracts were divided between two secondary reviewers, who independently screened their assigned portions. Discrepancies between the primary reviewer and the specific secondary reviewer who assessed the abstract were resolved through discussion. Similarly, this process was repeated for the full-text screening stage.

### Data collection process

All identified records were imported into EndNote [17] for organisation and removal of duplicates. Study selection was managed using Rayyan [18]. The primary reviewer developed and independently completed the data extraction form, with two secondary reviewers independently extracting data on assigned portions, guided initially by an example form to ensure consistency.

### Data items
We collected data on study design, participants, settings, AI interventions, human comparators, empathy measures, and key findings related to empathy.

### Synthesis methods
Despite heterogeneity, particularly in terms of empathy measurement instruments (ranging from custom Likert-type scales to qualitative coding) and empathy evaluators (patient proxies, medical students, etc.), studies were sufficiently similar for meta-analysis due to shared structural elements. All studies directly compared AI-generated responses to human HCPs using blinded evaluations (except one not stating blinding), and all but that same study quantitatively provided empathy results (mostly via single-item Likert scales), enabling relative effect size calculations. All comparisons were in text form. In one case, the text was converted to audio for patients, but empathy raters assessed only the written transcript. We also narratively synthesized the results (Appendix B). We organized studies by the large language model (GPT-3.5, GPT-4, other models) and then compared outcomes against its HCP comparators (e.g. physicians, nurses). Where available, we reported empathy scores, mean differences, p-values (for statistical significance), and other relevant summary statistics.

### Reporting bias assessment
Two reviewers independently assessed the risk of bias (ROB). The ROBINS-I tool [19] was used to assess the RoB across the 15 studies, given that 14 were non-randomized cross-sectional comparative analyses and one was a randomized controlled trial (RCT) [20]. Discrepancies were resolved through discussion, and with a third reviewer where necessary.

### Meta-analysis
We conducted a single meta-analysis with two subgroups (GPT-3.5 and GPT-4) due to their prevalence amongst the included studies (five and 10 appearances, respectively; other models appeared once). To prevent double-counting, we excluded additional AI arms from multi-model studies and entirely omitted studies evaluating both GPT versions on the same dataset. We extracted SMDs with 95% confidence intervals (CIs) and constructed random-effects models to pool effect sizes. Subgroup differences were examined by comparing pooled effect estimates, with $P < .05$ considered statistically significant. Analyses and plots used Review Manager (RevMan) 5.4 [21].

## Results
### Study selection
The search identified 987 unique results that were screened by title/abstract and by full text. Following title and abstract screening, 34 articles were included. During full-text screening, an additional 19 studies were excluded (Appendix C). Ultimately, 15 studies met the inclusion criteria for this systematic review (see Fig. 1).

### Risk of bias
Overall, nine studies had a moderate ROB, and six had a serious ROB (Appendix D). Seven used curated patient queries, potentially introducing selection bias. Four relied on Reddit communities [10, 22–24]—online forums where people publicly post health questions and receive free answers from strangers. These often attract users in 'desperate' circumstances [25], facing barriers to timely formal care, raising concerns about representativeness, thoroughness, and confounding. Other serious cases employed supervised designs (i.e. where a human expert reviewed AI outputs and blocked unsafe replies) [20, 26], complicating isolation of chatbot performance. Heterogeneity arose from diverse empathy raters (e.g. patient proxies, clinicians, psychology trainees), complicating comparisons because perceptions of empathy may vary by evaluator background. Finally, 14 of 15 studies assessed empathy using non-validated methods (e.g. single-item custom Likert scales), rather than a validated instrument (e.g. Consultation and Relational Empathy [CARE]) [27], limiting standardisation and heightening subjectivity.

### Characteristics of included studies
Fourteen of the 15 studies included in the review were published in 2024; the remaining study was published in 2023 [10].

#### Types of health concerns and specialty
The studies covered diverse health conditions and specialties (see Table 1). Four involved non-specific medical concerns, reflecting routine and general patient inquiries. These included outpatient queries across multiple departments [20], general health questions from social media [10, 28], and internal medicine patient portal interactions, including lab results and administrative requests [29]. Other studies addressed specialized clinical conditions, such as dermatology [26, 30], oncology [22], and thyroid conditions [31]. Chronic diseases were also covered, including systemic lupus erythematosus [32] and multiple sclerosis [27], alongside reconstructive surgery [33] and complete blood count lab result interpretation [23]. Two addressed mental and neurodevelopmental health, one examining mental health support inquiries [24] and the other focusing on autism-related queries [34]. Finally, one service-oriented study examined patient complaints from various departments [35].

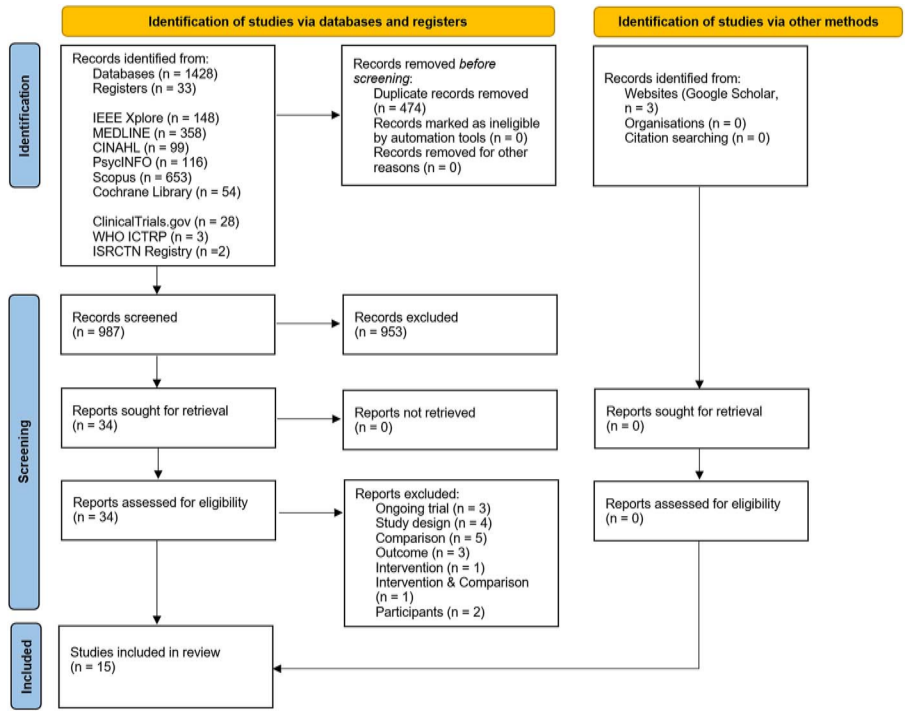

Figure 1. PRISMA flow diagram.

**Human HCP comparators**

The included studies compared AI-generated responses to a variety of human comparators. Two studies compared AI responses to those from surgeons [31, 33], while six studies involved comparisons with physicians specialising in specific fields [10, 22, 26, 27, 30, 34]. Three studies compared AI responses to physicians with no specified specialties [10, 23, 28]. Other comparisons included advanced practice providers in one study [33], nurses and frontline staff in one study [29], and reception nurses in another [20]. Mental health licensed professionals were the comparators in one study [24], whilst patient relations officers were included in another [35].

**Interaction modalities and AI models**

All but one of the included studies relied exclusively on text-based interaction with the AI system. In one study, patient speech was transcribed (through a real-time voice transcribing software) into text for the LLM (GPT-3.5) and then the AI's text response was converted back into audio [20], but the written transcript was ultimately used for the empathy evaluation. One study allowed patients to upload images, however they were described textually by a human intermediary before input [26]. All but one study [26] used versions of GPT (general-purpose language models). Three studies also evaluated GPT alongside other LLMs: one with ERNIE Bot [34], one with Claude [22], and one with Gemini Pro and Le Chat [23]. The exception was a study exclusively using Med-PaLM2, a model specifically designed for medical question-answering tasks [26].

**Empathy measurement tools**

All, bar one study, relied on unvalidated or custom tools for measuring empathy; the exception used the CARE scale [27], which is a validated instrument. Eight studies relied on single-item 1–5 Likert scales, where responses were rated from 'not empathetic' to 'very empathetic' or similar descriptors [10, 20, 24, 28, 30, 31, 33, 34]. One additional study used a 1–5 Likert scale, however, assessed separate scores for cognitive and emotional empathy [22]. Two studies used single-item 0/1–10 scales [32, 35] and another employed a single-item 1–6 scale [23]. Another study used a thematic coding framework, identifying and counting empathy-related statements—appreciation, acknowledgment, and compassion—as part of content analysis [26]. Additionally, one study measured empathy conditionally: reviewers first determined if responses (AI or human generated) were usable, then selected empathy as a reason for usefulness if applicable, rather than using any scale [29].

### Query sources

Seven studies utilized historical emails or message logs from private medical records [13, 27, 29–33, 35], including the smallest sample, four commonly asked patient emails [27]; six drew on publicly available questions from Reddit and other online forums [10, 22–24, 28, 34]; one collected real-time chat transcripts [26]; and one analysed in-person reception interactions, which yielded the largest sample, 2164 live outpatient queries [20]. Among the 11 studies that reported a specific location, most patient queries originated from Western countries. Four were from the USA [26, 29, 30, 33], one from Germany [28], one from Italy, one from Singapore [35], four from China [20, 31, 32, 34], and four from the global platform Reddit [10, 22–24], whose user base is predominantly from the USA and UK [36].

### How empathy was assessed

Across these studies, *all* evaluators functioned as 'observers' (not those asking/answering the inquiries) but varied in their backgrounds. One study used patient-proxies [27]; five employed HCPs [10, 20, 22, 29, 31]; three combined patient-proxies and HCPs [28, 31, 32]; three involved laypeople alongside HCPs [20, 30, 35]; one used medical students [33]; and one featured a psychology undergraduate and psychologist intern [24]. All reviewers in these studies were blinded to whether the responses were generated by AI or humans. The additional study relied on researchers or coders with mixed expertise in healthcare, conversational AI, and human-computer interaction, and blinding was not mentioned [26].

## Results of included studies

Table 1 summarizes the characteristics of the 15 included studies, with statistical results reported as presented in the studies (sub-scores, overall means, etc.). Detailed study-specific results, including granular metrics and model comparisons, are provided in Appendix E.

## Results of syntheses
### Overall summary of included comparisons

In 13 out of the 15 comparisons, one or more AI chatbots demonstrated a statistically significant advantage in projection of empathy over human healthcare practitioners [10, 20, 22–24, 27–29, 31–35] (see Fig. 2). In one of these 13 comparisons, one AI model (ERNIE Bot) of the two assessed did not show a statistically significant difference from humans, but the other model (GPT-4) within the same study did outperform ERNIE Bot, demonstrating a statistically significant result [34]. In the remaining two studies, both involving dermatology, human dermatologists outperformed AI (Med-PaLM 2 and ChatGPT-3.5) in perceived empathy [26, 30].

We first present the results of our meta-analyses for GPT-3.5 and GPT-4. Afterwards, we briefly discuss the two studies that could not be included in the meta-analysis (n = 1 to prevent double-counting [30], n = 1 due to missing outcome data [26]) and other LLM arms (GPT-3, Gemini, Le Chat, and ERNIE Bot) beyond GPT-3.5/4. These additional models were excluded from the meta-analysis to avoid overlapping data. A full narrative synthesis of all models and studies are included in Appendix B.

### Meta-analyses of GPT-3.5 and GPT-4 chatbots

Thirteen studies provided data suitable for meta-analysis. Overall, ChatGPT demonstrated significantly higher empathy than human practitioners (standardized mean difference [SMD] 0.87, 95% CI 0.54–1.20; $P < .00001$). Heterogeneity was moderate between GPT-3.5 and GPT-4 subgroups ($I^2 = 49.4\%$) but high across all studies ($I^2 = 97\%$). In Meyer et al. [23], the same 100 questions were tested with ChatGPT, Gemini, and Le Chat, risking double counting if pooled. Thus, it could not be pooled. As highlighted by Hussein, et al. [37], analyses of health record data are particularly prone to such double-counting errors. Similarly, Soroudi, et al. [33] compared GPT-3, GPT-4, and human HCPs; we retained only GPT-4 data to prevent double-counting. He, et al. [34] also tested ERNIE Bot and GPT-4 on the same dataset, so we included only GPT-4. Finally, Chen, et al. [22] simultaneously assessed GPT-3.5 and GPT-4 (and Claude) on the same dataset, creating an irresolvable conflict between our GPT-3.5 and GPT-4 subgroups; therefore, that study was omitted entirely from the pooled analysis. Li, et al. [26] evaluated Med-PaLM 2, but didn't provide specific metrics, so was similarly excluded from the following meta-analysis.

### Subgroup: GPT-3.5

Four studies provided data compatible with a GPT-3.5 subgroup meta-analysis (excluding Chen et al.). They reported empathy outcomes on validated (CARE) or custom Likert scales allowing calculation of standardized mean differences (SMDs). Pooled analysis (random-effects model) yielded an SMD of 0.51 (95%

Table 1. Empathy comparisons between AI chatbots and human healthcare practitioners—summary of findings across studies.

| Study | AI Model(s) | HCP Comparator(s) | Instrument/Scale or Approach (range of possible scores) | Evaluator Type | Reported Overall Difference (95% CI), *p*-value | Interpretation |
|---|---|---|---|---|---|---|
| **Armbruster (2024)** | ChatGPT-4 | Physicians (mixed specialties) | **Custom Likert** (1–5; 1 = very poor, 5 = very good). *Item: 'How empathetic or friendly is the response?'* | Patient-proxies and HCP observers | $P < .001$ | ChatGPT-4 statistically significantly outperformed the Expert Panel in empathy, as rated by both patients and specialists. |
| **Ayers (2023)** | ChatGPT-3.5 | Physicians (mixed specialties) | **Custom Likert** (1–5; 1 = not empathetic, 5 = very empathetic). *Item: 'Evaluate the empathy or bedside manner provided.'* | HCP observers | $P < .001$ | ChatGPT-3.5 responses were statistically significantly more empathic than physicians', with 45.1% of chatbot responses rated as empathic compared to only 4.6% for physicians. |
| **Chen (2024)** | GPT-3.5, GPT-4, Claude | Oncologists | **Custom Likert** (1–5; 1 = very poor, 5 = very good) assessing two dimensions—Cognitive and Emotional Empathy. *Item: 'Evaluate emotional and cognitive empathy.'* | HCP observers | $P < .001$ *(any chatbot vs. physicians)* | All three chatbots outperformed physicians (statistically significant). Claude had the highest empathy rating, followed by GPT-4, GPT-3.5, and then physicians. |
| **Guo (2024)** | ChatGPT-4 | Junior & Senior Surgeons | **Custom Likert** (1–5; 1 = very unsympathetic, 5 = very sympathetic). *Item: 'Evaluate the compassion of the response.'* | Patient-proxies and HCP observers | *Patient-reviewed comparison:* $P < .001$; *Surgeon-reviewed comparison:* $P = .007$ | ChatGPT-4's responses were rated statistically significantly more empathic than responses from both junior and senior specialists. The difference was larger (in statistical terms) for patient reviewers than for surgeon reviewers. |
| **He (2024)** | ChatGPT-4, ERNIE Bot | Physicians (mixed specialties) | **Custom Likert** (1–5; 1 = lacking, 5 = very humane). *Item: 'Evaluate empathy in terms of respect, communication, compassion, and emotional connection.'* | HCP observers | *Physicians vs. ChatGPT:* $P < .001$; *Physicians vs. ERNIE Bot:* $P = .14$; *ChatGPT vs. ERNIE Bot:* $P < .001$ | ChatGPT-4 scored significantly better than physicians. ERNIE Bot scored slightly lower than physicians, but not significantly so. |
| **Li (2024)** | Med-PaLM 2 | Dermatologists | **Frequency count of empathy markers** ('appreciation', 'acknowledgment', 'compassion') via qualitative coding of transcripts. | Coders (observers) with mixed expertise in artificial intelligence/healthcare/human-computer interaction | $P < .05$ *(between groups across appreciation, acknowledgment, compassion markers)* | Human clinicians used more total empathy-related language (across 'appreciation', 'acknowledgment', 'compassion' markers)—this difference was statistically significant—but AI used more 'compassion' markers. |
| **Maida (2024)** | ChatGPT-3.5 | Neurologists | **CARE scale** (overall 10–50; 10 items rated on a 5-point Likert [1 = poor, 5 = excellent]). *Measures therapeutic empathy during one-on-one consultations.* | Patient proxies | 1.38 (0.65–2.11); $P < .01$ | ChatGPT-3.5 was rated higher in empathy than neurologists, with a statistically significant overall difference. |

Table 1. Continued.

| Study | AI Model(s) | HCP Comparator(s) | Instrument/Scale or Approach (range of possible scores) | Evaluator Type | Reported Overall Difference (95% CI), *p*-value | Interpretation |
|---|---|---|---|---|---|---|
| **Meyer (2024)** | ChatGPT-4, Gemini Pro, Le Chat | Physicians (mixed specialties) | **Custom Likert** (1–6; 1 = excellent, 6 = inadequate). 'Level of empathy' with response distributions provided as percentages across categories. | HCP observers | *ChatGPT vs. Physicians*: P < .001; *Gemini vs. Physicians*: P < .001; *Le Chat vs. Physicians*: P < .001; *ChatGPT vs. Gemini*: P = .50; *ChatGPT vs. Le Chat*: P < .001; *Gemini vs. Le Chat*: P < .001 | All three chatbots were rated as more empathic than physicians. Among chatbots, ChatGPT and Gemini did not differ significantly; both were rated statistically significantly higher than Le Chat. |
| **Reynolds (2024)** | ChatGPT-3.5 | Dermatologists | **Custom Likert** (1–5; 1 = very poor, 5 = very good). Item: '*Evaluate the level of empathy demonstrated*.' | Laypeople observers and HCP observers | *Physician raters*: P = .001; *Non-Physician raters*: P = .09 | Physicians were rated as more empathic by physician reviewers (significant). Non-physicians also tended to rate physicians higher, but that was not statistically significant. |
| **Small (2024)** | ChatGPT-4 | Variety of HCPs (physicians, nurses, and frontline staff) | **Count of empathic responses** (i.e. responses classified as 'empathic') plus linguistic analysis (measures of subjectivity and positive polarity). Item: '*Would you find this draft useful? (Empathy options)*' | HCP observers | (No single 'overall difference' beyond the above proportions.) | ChatGPT-4 generated a higher proportion of empathic responses than healthcare professionals; its responses featured statistically significantly greater subjectivity and positive language. |
| **Soroudi (2024)** | ChatGPT-3/4 (Full vs. Brief) | Plastic Surgeons & Advanced-Practice Providers | **Custom Likert** (1–5; 1 = not empathetic, 5 = very empathetic). Item: *Rate empathy for both full and brief response formats.* | Medical Students (observers) | *Combined Chatbot vs. Providers*: P < .001; *Brief Chatbot vs. Providers*: P = .125 | ChatGPT-generated responses (any version) were perceived as statistically significantly more empathic than plastic surgeon/APP responses overall. The brief ChatGPT responses alone did not differ significantly from providers' scores, but full ChatGPT responses did. |
| **Wan (2024)** | 'SSPEC' (GPT-3.5 with supervision) | Reception nurse-only | **Custom Likert** (1–5; 1 = detached tone, 5 = strongly connects). Item: '*Empathy— consideration for the patient's perspective.*' | Laypeople observers and HCP observers | P < .001 (Internal Validation) P < .001 (RCT) | The SPPEC scored statistically significantly higher in empathy than nurses in both internal validation (retrospective) and RCT (prospective). |
| **Xu (2024)** | ChatGPT-4 | Rheumatologists | **Custom Likert** (0–10); 0 = no empathy and 10 = highly empathetic. Item: '*Evaluate how well the answer conveys understanding and care for patient concerns.*' | Patient-proxies and HCP observers | **Overall (Chinese + English)**: *Rheumatologists' difference*: 0.46 (0.23–0.69); P < .01 **Chinese only**: *Rheumatol. Eval*: 0.48; P < .001 *SLE Patients Eval*: 0.18; P = .249 **English only**: *Rheumatol. Eval*: 0.13; P = .80 *SLE Patients Eval*: 1.32; P < .001 | ChatGPT-4 outperformed rheumatologists overall with a statistically significant difference. For Chinese evaluations, both rheumatologists and patients rated ChatGPT-4 higher (with statistical significance for rheumatologists only), while in English, only patient ratings were statistically significantly higher. |

Table 1. Continued.

| Study | AI Model(s) | HCP Comparator(s) | Instrument/Scale or Approach (range of possible scores) | Evaluator Type | Reported Overall Difference (95% CI), *p*-value | Interpretation |
|---|---|---|---|---|---|---|
| Yonatan-Leus (2024) | ChatGPT-4 | Mental Health Licensed Professionals | **Custom Likert** (1–5; 1 = no empathic concern, 5 = strong empathic concern). Item: *'Evaluate empathic concern.'* | Psychology undergraduate and psychologist intern (both observers) | 0.60 (large effect); *P* < .001 | ChatGPT-4 was perceived as statistically significantly more empathic than human mental health professionals, consistently showing greater warmth—a statistically significant difference with a large effect size. |
| Yong (2024) | ChatGPT-4 | Patient Relations Officers | **Custom Likert** (1–10; 1 = not empathetic, 10 = very empathetic; no midpoint provided). Item: *'Empathy.'* | Laypeople observers and HCP observers | *P* < .001 | ChatGPT-4 was rated higher than humans in empathy. This difference was statistically significant. |

CI -0.14–1.16) favouring GPT-3.5 over human healthcare practitioners ($I^2 = 99\%$), indicating high heterogeneity. This result was not statistically significant, however ($P = .12$), due to the one instance where HCPs scored higher in Reynolds, et al. [30].

Across four studies, GPT-3.5 outperformed human clinicians in three settings. Maida, et al. [27] reported an SMD of 0.15 (95% CI: 0.07–0.23) for neurologic inquiries, Wan, et al. [20] found an SMD of 0.79 (95% CI: 0.70–0.87) in an outpatient reception setting, and Ayers, et al. [10] observed an SMD of 1.91 (95% CI: 1.67–2.15) based on social media queries. In contrast, Reynolds, et al. [30] showed that for dermatology queries, human responses were rated higher (SMD −0.99, 95% CI: −1.52 to −0.46). We narratively synthesize these results in Appendix B.

### Subgroup: GPT-4

Nine studies contributed data for the GPT-4 subgroup, yielding a pooled SMD of 1.03 (95% CI 0.71–1.35) in favour of GPT-4 ($I^2 = 87\%$, Tau [2] = 0.19), again reflecting high heterogeneity. This result was statistically significant ($P < .00001$).

Across nine studies, GPT-4 consistently outperformed human clinicians. Guo, et al. [31] reported an SMD of 1.42 (95% CI: 0.85–1.99) for thyroid-related inquiries; Yonatan-Leus and Brukner [24] found an SMD of 0.97 (95% CI: 0.73–1.21) for mental health queries on social media; and Soroudi, et al. [33] observed an SMD of 0.83 (95% CI: −0.09–1.75) for breast reconstruction questions. Armbruster, et al. [28] showed an SMD of 1.44 (95% CI: 1.13–1.76) in a web-based setting, while He, et al. [34] reported an SMD of 0.80 (95% CI: 0.62–0.99) for autism-related inquiries. Xu, et al. [32] found an SMD of 0.23 (95% CI: −0.06–0.51) for systemic lupus erythematosus questions, and Yong, et al. [35] reported an SMD of 2.08 (95% CI: 1.27–2.88) for patient complaints. Additionally, Small, et al. [29] observed an SMD of 0.48 (95% CI: 0.16–0.80) for outpatient internal medicine queries, and Meyer, et al. [23] reported an SMD of 1.44 (95% CI: 1.12–1.75) for laboratory-interpretation queries. We narratively synthesize these results in Appendix B.

### Subgroup comparison: GPT-3.5 vs. GPT-4

When examining GPT-3.5 and GPT-4 side by side, GPT-4 consistently outperformed human clinicians in empathy, whilst GPT-3.5 showed mixed results and failed to demonstrate a statistically significant advantage. A subgroup analysis testing for differences between GPT-3.5 and GPT-4 revealed no statistically significant difference ($P = .16$). Thus, while GPT-4 achieved more consistent results, the current data do not conclusively indicate that it is definitively more empathic than GPT-3.5 (Fig. 2). The pooled SMD of both models (n = 13 pooled studies) was 0.87 (95% CI: 0.54–1.20), with both GPT-3.5 and GPT-4 collectively demonstrating statistically significantly ($P < .00001$) higher empathy ratings than HCPs.

### Non-meta-analysed results

Across the non-meta-analysed studies, Chen, et al. [22] found that GPT-3.5, GPT-4, and Claude all outperformed oncologists on oncology queries—with GPT-4 scoring highest. Li, et al. [26] showed that, while human clinicians used more empathy-related language overall, Med-PaLM 2 demonstrated a slight advantage in expressing compassion. Meyer, et al. [23] reported that both Gemini Pro and Le Chat (Mistral Large) were rated significantly higher than physicians, whereas He, et al. [34] found that ERNIE Bot slightly underperformed compared to clinicians.

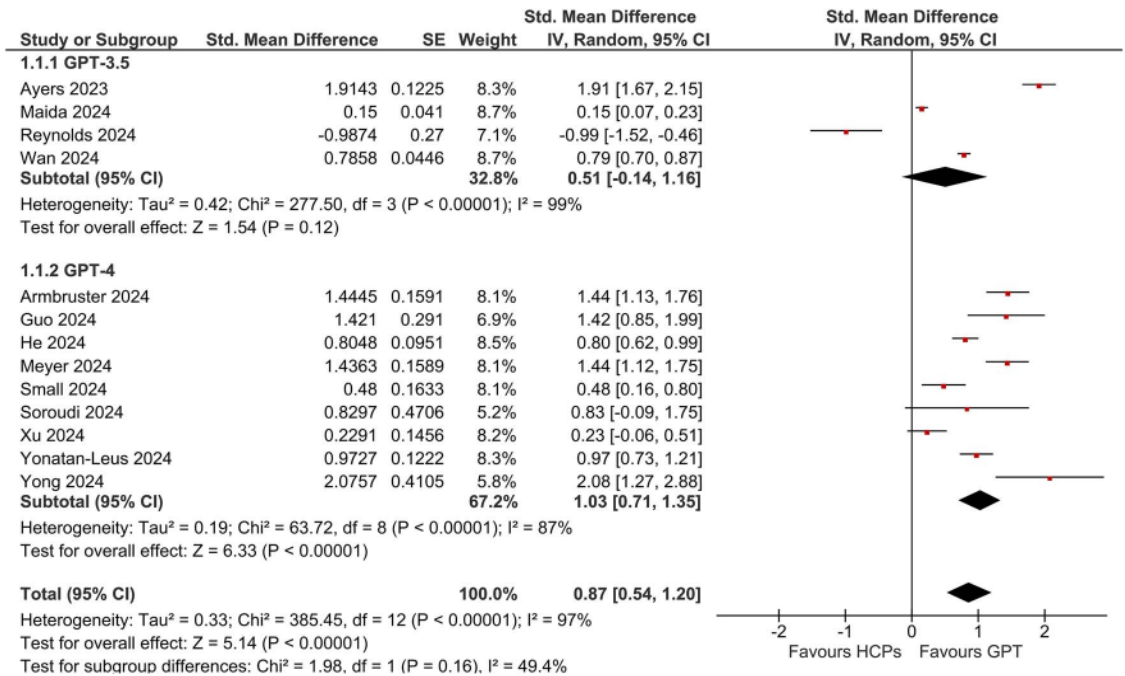

Figure 2. Forest plot comparing empathy ratings: GPT-3.5 vs. human practitioners and GPT-4 vs. human practitioners (excluding overlap).

Finally, Soroudi, et al. [33] observed that GPT-3's 'full' format outperformed its 'brief' format, with both formats exceeding providers' ratings. For full details, please refer to Appendix B.

## Discussion

### Interpretation of results

This is the first systematic review, we are aware of, that compares the empathy of AI chatbots with human practitioners. Our meta-analysis of 13 studies shows that ChatGPT has a 73% likelihood of being perceived as more empathic than a human practitioner in a head-to-head matchup, using text-based interactions (representing the probability of superiority). This finding was across multiple clinical specialties and various evaluative methods. This would be roughly equivalent to a difference of 2 points on a 10-point scale. An important implication is that text-based AI-driven interactions are unlikely to cause harm through deficits in empathy, aligning with broader evidence supporting AI's potential to enhance healthcare engagement, quality, and efficiency [3].

### Limitations

This study has several limitations. Firstly, all studies analysed text-based interactions. This is a problem because empathy in healthcare consultations often relies on both verbal and non-verbal cues (e.g. nodding and leaning forward) [38], and because text-based communication represents a relatively small portion of healthcare interactions [39]. That being said, healthcare practitioners are increasingly using purely text-based communication, suggesting that, at least in these cases, our results are relevant to actual practice [40]. Another limitation is that the included studies evaluated empathy from proxy measures rather than the perspectives of the patients directly receiving care. Given that HCP and direct care recipients' empathy ratings have been shown to differ [41, 42], it is possible that patient ratings would have been different. Also, only two of the 15 [29, 30] involved HCPs replying to their own patients with access to records and prior care context; the remaining studies assessed one-off interactions, often from public forums such as Reddit, where tone and disclosure may differ from private clinical settings.

Additionally, the potential gains in empathic communication must be weighed against ongoing concerns regarding the reliability of AI-driven clinical content [43]; any benefits in empathic delivery risk being overshadowed if the medical advice offered is inaccurate. Although empathy in healthcare (also called therapeutic or clinical empathy) is widely recognized as the ability to understand a patient's emotions [44], the included studies defined and operationalized it in varied ways, with some using wording that overlapped with related constructs such as compassion, complicating comparability and interpretability.

Moreover, six of the included studies sourced patient questions and/or clinician responses from public forums. It could be that the AI chatbots (trained on vast internet datasets) [45] encountered the material during training, giving them an unfair advantage through test-set contamination. However, given the enormous volume of training data, any influence from specific queries is likely minimal. Also, models typically generate original, context-specific responses rather than reproducing exact copies [45]. However, this limitation could also work against the success of the AI chatbots: since human responses were generally less empathic, mimicking them might reduce the chatbots' relative empathy scores. Most (14/15) of the included studies focused on GPT-3/3.5 or GPT-4, which may limit generalisability to clinical practice. Widely deployed clinical chatbots (e.g. Wysa) often rely on proprietary models with distinct architectures and training data that could influence empathic communication. This variability raises questions about how well the findings translate to tools used in real-world care. Finally, all included studies were from 2024/2023; newer models such as GPT-4.5, released in February 2025 and claimed by OpenAI to demonstrate greater emotional intelligence [46], and GPT-5 have not yet been evaluated in the literature. Nine of 15 studies lacked a priori power calculations and used convenience samples, limiting precision and generalisability. Most outcome measures were unvalidated (e.g. single-item Likert scales, proportion-based assessments), with only one study using the validated CARE scale [27] and three reporting reliability testing [24, 32, 34], reducing comparability. Comparator groups varied widely in roles and specialties, so the pooled average offers a broad estimate but should not be over-interpreted for specific roles, and no study compared empathy across them (e.g. surgeons vs. mental health professionals).

In addition, the evaluators of empathy varied (including patient proxies, lay people, students, and HCPs, often in mixed combinations), further complicating pooled inferences. Relatedly, many studies lacked the detailed statistics (e.g. means, standard deviations, or standard errors) typically required for SMD calculations, necessitating approximations or the combination of categories following guidance in the Cochrane Handbook [47].

Finally, while statistically significant empathy differences were identified, their clinical relevance—such as direct impacts on patient outcomes—remains uncertain. However, the magnitude of these differences (∼20% absolute difference) suggests potential clinical relevance, warranting further investigation into how enhanced AI empathy might influence healthcare outcomes.

## Recommendations for future research and practice

Future research should explore leveraging AI's empathic advantages in text-based communication without compromising accuracy or safety. One promising approach is using AI to draft patient-facing messages, such as responses to medical or administrative queries, freeing clinician time and enhancing care quality [10]. Building on this, we propose a collaborative human–AI interaction model: clinicians produce an initial response, while AI augments these drafts by refining tone and integrating empathy, functioning as an 'empathic enhancer'. This approach, ensuring accuracy through clinician oversight (as advocated by Reynolds, et al. [30] and Chen, et al. [22]), mitigates risks of AI-generated inaccuracies by having clinicians generate the core content, with AI solely providing refinement. Rigorous randomized trials are recommended to evaluate impacts on patient satisfaction and clinician workload.

Emerging voice-enabled chatbots (e.g. ChatGPT's Advanced Voice Mode) claim capabilities to 'respond with emotion' and 'pick up on non-verbal cues' [48], yet no studies in our review compared these systems to HCPs. Given the observed empathic advantage of AI in text-only interactions, trials in telephone consultations (which account for 26% of all GP appointments [49]) could test whether this persists in voice communications.

Nearly all studies blinded raters to the source of each response (AI vs. human) to mitigate bias. However, real-world standards require disclosing AI involvement. This disclosure might have diminished perceived empathy once evaluators know the response is AI-generated. Future studies should investigate empathy ratings in scenarios where participants/reviewers are explicitly informed if they are communicating with AI, to ascertain whether the favourable impressions seen under blinded conditions persist. Supporting this concern, Perry et al. [50] found that AI-assisted replies were initially rated more empathic than human ones in online emotional-support chats, but this advantage disappeared once users learned the responses came from AI.

Further research is necessary on how prompt design influences empathic outcomes. For instance, two studies highlighted that restricting AI's response length—matching typical clinician brevity—reduces perceived empathy, while allowing longer responses correlates with higher ratings [10, 33]. Additionally, studies did not instruct the chatbot to emphasize empathy, potentially influencing the outcomes, with the study investigating Med-PaLM 2 suggesting it fell short in comparison to humans, as the 'agent was not explicitly prompted to produce empathic language' [26]. Optimising prompts to balance accuracy, brevity, and empathy is essential for real-world implementation. Beyond prompt engineering, empathy benchmarking should expand to diverse models beyond the GPT family (e.g. Claude, Llama) to identify model-specific strengths across clinical contexts.

Finally, further research should assess patients' perceptions of empathy based on their own experiences of healthcare consultations.

## Conclusion

Our review indicates that generative AI chatbots—particularly GPT-4—are often perceived as more empathic than human practitioners in text-based interactions, a finding consistent across various clinical contexts though with notable exceptions in dermatology. While methodological limitations, such as reliance on unvalidated scales and text-only evaluations, temper these results, the compelling evidence challenges longstanding assumptions about human clinicians' exclusive capacity for empathic communication—including assertions from the 2019 Topol Review (a UK government-commissioned roadmap for healthcare technology), which deemed 'empathy and compassion' an 'essential human skill[s] that AI cannot replicate' [8]. Future research should extend to evaluations using voice-based interactions and direct patient feedback, and ensure rigorous validation and transparency through randomized trials to uphold clinical reliability.

## Acknowledgements

We thank Selina Lock, Research Librarian, for verifying the search strategy.

## Author contributions

Alastair Howcroft (Conceptualization, Data curation, Formal analysis, Investigation, Methodology, Project administration, Software, Validation, Visualization, Writing—original draft, Writing—review & editing), Amber Bennett Weston (Supervision, Writing—review & editing), Ahmad Khan (Data curation), Joseff Griffiths (Data curation), Simon Gay (Writing—review & editing), and Jeremy Howick (Supervision, Writing—review & editing)

## Supplementary data

Supplementary data are available at *British Medical Bulletin Journal* online.

## Conflict of interest

The authors declare that they have no known competing financial interests or personal relationships that could have appeared to influence the work reported in this paper.

## Funding

This research did not receive any specific grant from funding agencies in the public, commercial, or not-for-profit sectors.

## Data availability

All data underlying this article are available within the main text and its supplementary materials (Appendices A–E), which include the full search strategy, list of excluded studies, narrative syntheses, ROB assessments, and study-specific results. No additional datasets were generated or analysed.

## Protocol and registration

The protocol was developed following the Preferred Reporting Items for Systematic Reviews and Meta-Analysis Protocols (PRISMA-P) guidelines and was registered prospectively with the Open Science Framework on 28 October 2024 (https://osf.io/2rt3f). Contrary to the original protocol's description of a scoping review, this study was conducted as a systematic review with meta-analysis. Following protocol registration, clarifications were incorporated to refine the eligibility criteria. These included specifying the inclusion of studies utilising patient-generated data (e.g. emails), defining 'AI' as LLMs to improve scope precision, and broadening 'healthcare settings' to encompass informal health-related online contexts. In response to the observed quantitative outcome measures across included studies, the analysis plan was adapted from the originally proposed narrative synthesis to a meta-analysis, enabling the estimation of an overall effect size.